\documentclass[12pt,preprint]{aastex}

 \usepackage{graphics}
 \usepackage{graphicx}
 \usepackage{epsfig}

\usepackage{amssymb}
\usepackage{amsmath}
\usepackage{aas_macros}

\slugcomment{Astrophysics \& Space Science, doi:10.1007/s10509-011-0632-y}

\shorttitle{Trajectories of $L_4$ and Lyapunov Characteristic Exponents(LCEs) \dots Chermnykh-Like Problem}
\shortauthors{Badam Singh Kushvah}

\begin{document}

\title{Trajectories of $L_4$ and Lyapunov Characteristic Exponents in the Generalized Photogravitational Chermnykh-Like Problem}


\author{Badam Singh Kushvah}
\affil{Department of   Applied  Mathematics,
Indian School of Mines, Dhanbad - 826004\\ Jharkhand(India)}
\email{bskush@gmail.com,kushvah.bs.am@ismdhanbad.ac.in}



\begin{abstract}
 The dynamical  behaviour of near by trajectories is being  estimated by Lyapunov Characteristic Exponents(LCEs)  in the Generalized Photogravitational Chermnykh-Like problem. It is found that the  trajectories of the Lagrangian  point $L_4$ move along the  epicycloid path, and   spirally depart from the vicinity of the point. The LCEs remain positive for all the cases and   depend on the initial deviation vector as well as renormalization time step. It is noticed that the trajectories  are chaotic in nature and  the $L_4$ is  asymptotically stable. The   effects of radiation pressure, oblateness and  mass of the belt  are also examined in the present model. 
\end{abstract}





\section{Introduction}
\label{sec:Int}
In present paper our aim is  to obtain trajectories of $L_4$ and  is to estimate the rate of deviation for initially closely related trajectories  in  the modified  restricted
three body problem model(as in \cite{Kushvah2008Ap&SS.315,Kushvah2009Ap&SS}) with  radiation from  Sun, oblateness of the second primary(massive body)  and influence of the belt. It is supposed that 
the  primary bodies  and a belt  are moving in a circular orbits about the common  center of mass of both primaries.  First time such problem  was discussed by  \citet{Chermnykh1987} and its  importance in
astronomy has been addressed  by \cite{Jiang2004IJBC}.  More generalized  cases of the problem were studied by
many scientists such as  \cite{JiangYeh2004AJ}, \cite{Papadakis2004A&A},\cite{Papadakis2005Ap&SS299} and \cite{JiangYeh2006Ap&SSI,Yeh2006Ap&SS.306..189Y}. The effect of radiation pressure, Poynting-Robertson(P-R)drag and oblateness  on the  linear stability and nonlinear stability of the $L_{4(5)}$ have been  discussed by  \cite{KushvahBR2006} ; \cite{Kushvah2007BASI,Kushvah2007Ap&SS.312,Kushvah2007EM&P}. In our  article  \cite{Kushvah2010Ap&SS.tmp..286K}, we have   described the  design of the trajectory and analysis of the stability of collinear point $L_2$ in the Sun-Earth system.

The first fundamental article about LCN's was written by \cite{Oseledec1968} in their study of the ergodic theory of dynamical system and \cite{Benettinetal1980Mecc...15....9B}  presented  explicit methods for computing all LCEs of a dynamical system. Then \cite{Jefferys1983CeMec..30...85J} examined stability in the restricted problem of three bodies with Liapunov Characteristic number. First time \cite{Wolf1985285} presented an algorithm with  FORTRAN code that allows to estimate non-negative Lyapunov Exponents(LEs) from an experimental time series. \cite{MSandri} have presented method for numerical calculation of Lyapunov Exponents for a smooth dynamical system with    Mathematica[\cite{Wolfram2003}]  code.  \cite{Tancredi2001AJ....121.1171T}  compared two different methods to compute Lyapunov Exponents(LEs). They have shown that since the errors are introduced in the renormalization procedure, it is natural to expect a dependency of the estimated LCEs with the number of renormalization performed in the sense that the smaller the step the worse the estimation. In his study they made conclusion  that the two-particle method is not recommended to calculate LCEs in these cases where the solution can fall in a region of regular or quasi regular solution of the phase space. For a region of strong stochastically  the LCEs calculated with the two-particle method gives acceptable value. 

This paper is organized as follows: In section \ref{sec:TrjL4}, we state the model of the dynamical system and compute the trajectories of $L_4$. Section \ref{sec:lces} gives method to compute the LCEs, where subsection \ref{subsec:1lce} presents the first order  LCEs for various set values of parameters, time ranges  and renormalization time steps. Section \ref{sec:stbL4} presents comment about stability using trajectories of $L_4$. Lastly, section \ref{sec:con} concludes the paper. 

\section{Trajectory  of  $L_4$}
\label{sec:TrjL4}
It is  supposed that the motion of an  infinitesimal mass particle be  influenced by the gravitational force from  the two primaries(massive bodies) and a belt of mass   $M_b$.  We also assume that infinitesimal mass does not influence the motion of the two massive bodies which move  in circular orbit under their mutual gravitational attraction.  Let us assume that $m_1$ and  $m_{2}$   be the masses of the bigger and  smaller primary  respectively, $m$ be the mass of the infinitesimal body. The units  are normalized by supposing  that the sum of the  masses  to be unity, the distance between both massive bodies to be unity. The rotating frame normalized to rotate with unit angular velocity   and  the time is normalized in such a way that  the time  for one period as a unit so that,  the Gaussian constant of gravitational $\Bbbk^{2}=1$. For the present  model,  perturbed mean motion $n$ of the primaries is given by  $n^{2}=1+\frac{3A_{2}}{2}+\frac{2M_b r_c}{\left(r_c^2+T^2\right)^{3/2}}$, where $T=\mathbf{a}+\mathbf{b}$, $\mathbf{a,b}$ are flatness and  core parameters respectively[as in \cite{Yeh2006Ap&SS.306..189Y}] which determine the density profile of the belt; where  $r_c^2=(1-\mu)q_1^{2/3}+\mu^2$,   $A_{2}=\frac{r^{2}_{e}-r^{2}_{p}}{5r^{2}}$ is the oblateness coefficient of $m_{2}$; $r_{e}$, $r_{p}$  are the equatorial and polar radii of $m_{2}$ respectively,  $r$ is  the distance between primaries and the radius of the belt;  $\mu=\frac{m_{2}}{m_{1}+m_{2}}$ is a  mass   parameter;   $q_1=1-\frac{F_p}{F_g}$ is a mass reduction factor and   $F_{p}$ is the solar radiation pressure force  which is exactly apposite to the gravitational attraction force $F_g$. In a rotating reference  frame   the coordinates of $m_1$ and   $m_2$ are  $(-\mu,0)$ and  $(1-\mu,0)$  respectively.
We consider the model proposed by \cite{MiyamotoNagai1975PASJ}, and  equations of motion are given as in  \cite{Kushvah2008Ap&SS.315} and \cite{Kushvah2009Ap&SS}:

\begin{eqnarray}
\ddot{x}-2n\dot{y}&=&\Omega_x ,\label{eq:Omegax}\\
\ddot{y}+2n\dot{x}&=&\Omega_y,\label{eq:Omegay}
 \end{eqnarray}
where
\begin{eqnarray*}
\Omega_x&=& n^{2}x-\frac{(1-\mu)q_1(x+\mu)}{r^3_1}-\frac{\mu(x+\mu-1)}{r^3_2}-\frac{3}{2}\frac{\mu{A_2}(x+\mu-1)}{r^5_2}\nonumber\\
&&-\frac{M_b x}{\left(r^2+T^2\right)^{3/2}} -\frac{W_1}{r^2_1}\biggl[\frac{(x+\mu)}{r^2_1}\{(x+\mu){\dot{x}+y\dot{y}}\} +\dot{x}-ny \biggr],\\
\Omega_y&=&n^{2}y
-\frac{(1-\mu)q_1{y}}{r^3_1}
-\frac{\mu{y}}{r^3_2}-\frac{3}{2}\frac{\mu{A_2}y}{r^5_2}\nonumber\\&&-\frac{M_b y}{\left(r^2+T^2\right)^{3/2}}-\frac{W_1}{r^2_1}\biggl[\frac{y}{r^2_1}\{(x+\mu)\dot{x}+y\dot{y}\}+\dot{y}+n(x+\mu)\biggr],\end{eqnarray*}
\begin{eqnarray*}
\Omega&=&\frac{n^2(x^2+y^2)}{2}+\frac{(1-\mu)q_1}{r_1}+\frac{\mu}{r_2}+\frac{\mu
 A_2}{2r_2^3}+\frac{M_b}{\left(r^2+T^2\right)^{1/2}}\nonumber\\&&+W_1\left\{ \frac{(x+\mu)\dot x + y\dot y}{2r_1^2}-n \arctan\left( \frac{y}{x+\mu}\right)\right\},\\
W_1&=&\frac{(1-\mu)(1-q_1)}{c_d},\ r_1^2=(x+\mu)^2+y^2,\ r_2^2=(x+\mu-1)^2+y^2.   
\end{eqnarray*}

The parameter $W_1$  is  considered due to P-R drag[more review in  \cite{Poynting1903},\cite{Robertson137}, \cite{Chernikov1970}, \cite{Murray1994} and \cite{Kushvah2009RAA}]. Where  $r_1$, $r_2$ are the distances of $m$ from first and second primary respectively.  The  dimensionless velocity of the light is supposed to be  $c_d= 299792458$.  Then from  equations (\ref{eq:Omegax}) and (\ref{eq:Omegay}) energy integral is given as:
\begin{equation}
E=\frac{1}{2}\left(\dot{x}^2+\dot{y}^2\right)-\Omega(x,y,\dot{x},\dot{y})=(\mbox{Constant})\label{eq:E}\end{equation}
 where the quantity $E$ is an energy integral related to the Jacobi's constant $C(=-2E)$.

\begin{figure}
\begin{center}
\includegraphics[scale=.5]{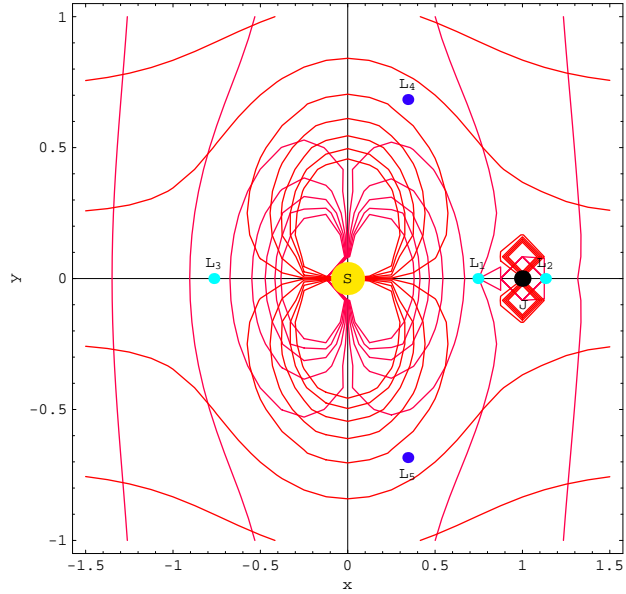}
\end{center}
\caption{The position of equilibrium points.}\label{fig:lpts}
\end{figure}

For numerical computation of equilibrium points, we  divide the  orbital plane  $Oxy$  into three parts  $x\leq-\mu$,  $-\mu<x<1-\mu$ and $1-\mu\leq x$ with respect to the primaries. For the   simplicity,  we set  $\mu= 9.537 \times 10^{-4}, T=0.01$.  The equilibrium points are given by  substituting $\Omega_x=\Omega_y=0$,  and   presented  in figure
 \ref{fig:lpts} when $q_1=0.75$, $A_2=0.25$, $M_b= 0.25$.  In this figure the  dark blue dotes present the position of  $L_4(5):(x=0.347988,y=\pm 0.70645)$, the light blue represent  the collinear equilibrium points $L_1:x=0.753578,L_2:x=1.14795$ and $L_3:x=-0.788385$ for which $y=0$.

 The equations (\ref{eq:Omegax}-\ref{eq:Omegay})
with initial conditions $x(0)=\frac{1-2\mu}{2}$, $y(0)=\frac{\sqrt{3}}{2}$, $x'(0)=y'(0)=0$ are
used to determine the trajectories of $L_4$ for different possible
cases. At at time $t=0$, the origin of coordinate axes  is supposed at the equilibrium
point. 
\begin{figure}
\begin{center}
\includegraphics[scale=.65]{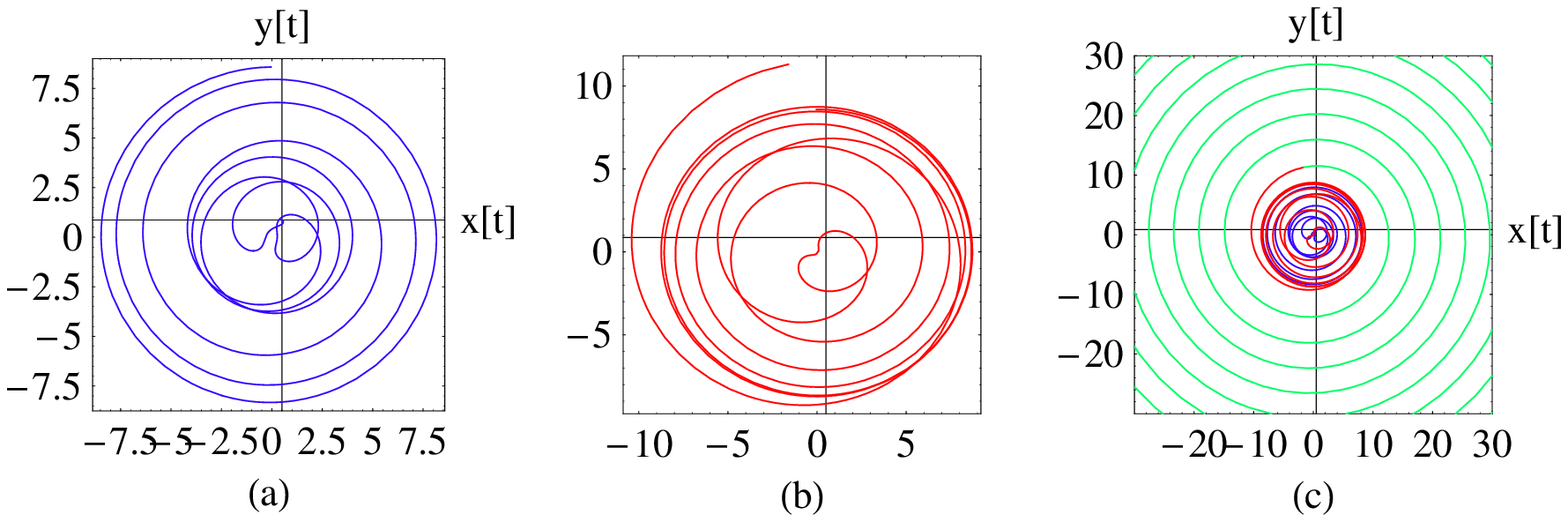}\\
\includegraphics[scale=.65]{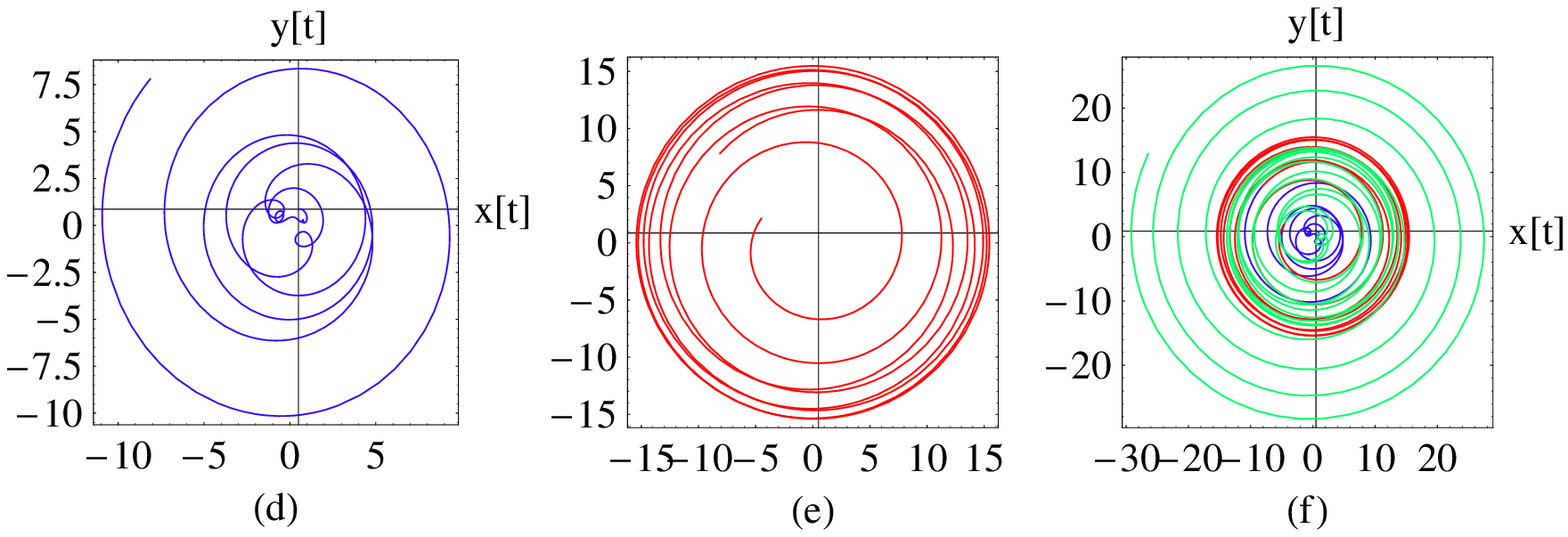}
\end{center}
\caption{Trajectory of $L_4$ when $q_1=0.75,A_2=0.0$ in frame(a)  blue curve  for $0 \leq t \leq 50$, (b) red for $50 \leq t \leq 100$ and (c) green   for $0 \leq t \leq 200$. Frames(a-c) $M_b=0.0$, and frames (d-f) $M_b=0.25$.}\label{fig:trjq75ma0}
\end{figure}

In the present model all the computed trajectories of the  $L_4$ follow  approximately the same path described by an epitrochoid  whose parametric equations are given as:
\begin{eqnarray}
    x (t)& =& (a_1 + b_1) \cos t - d_1 \cos \left( \frac{a_1 + b_1}{b_1} t \right)\label{eq:epitx}\\
    y (t)& = &(a_1 + b_1) \sin t - d_1 \sin \left( \frac{a_1 + b_1}{b_1} t \right)\label{eq:epity}
\end{eqnarray}

where $a_1$ is radius of a fixed circle, $b_1$ is radius of rolling circle and $d_1$ is distance form center of rolling circle to to the point$(x(t),y(t))$ which forms a trajectory. It is evident from above equations that if $d_1$ depends on  time then orbit is unstable and trajectory moves  spirally outward the vicinity of the initial point.

When $q_1=0.75,A_2=0$, the trajectory is shown in figure \ref{fig:trjq75ma0} with panels(a-d) for $M_b=0.0$ and panels(e-f) for $M_b=0.0$, where frames(a\& d) $0\leq t \leq 50$, (b\&e) $50 \leq t \leq 100$ and (c\&f) $0 \leq t \leq 200$.  It is  clear from figure that  if $0 \leq t \leq 50$ and  $M_b=0.0$ the trajectory of $L_4$ is similar to the curve described by  epitrochoid (\ref{eq:epitx}, \ref{eq:epity}) for $a_1=1/7,b_1=1=d_1$. When  $t>50$ then  $d_1$  becomes function of time $t$, and the trajectory moves  spirally outward.  When $M_b=0.25$, the trajectory follows the path correspond to parameters $a_1=1,b_1=\sqrt{5}$(irrational), $d_1=3$.    Here the value of $b_1$ is irrational number which shows that the motion is non periodic. 
\begin{figure}
\begin{center}
\includegraphics[scale=.6]{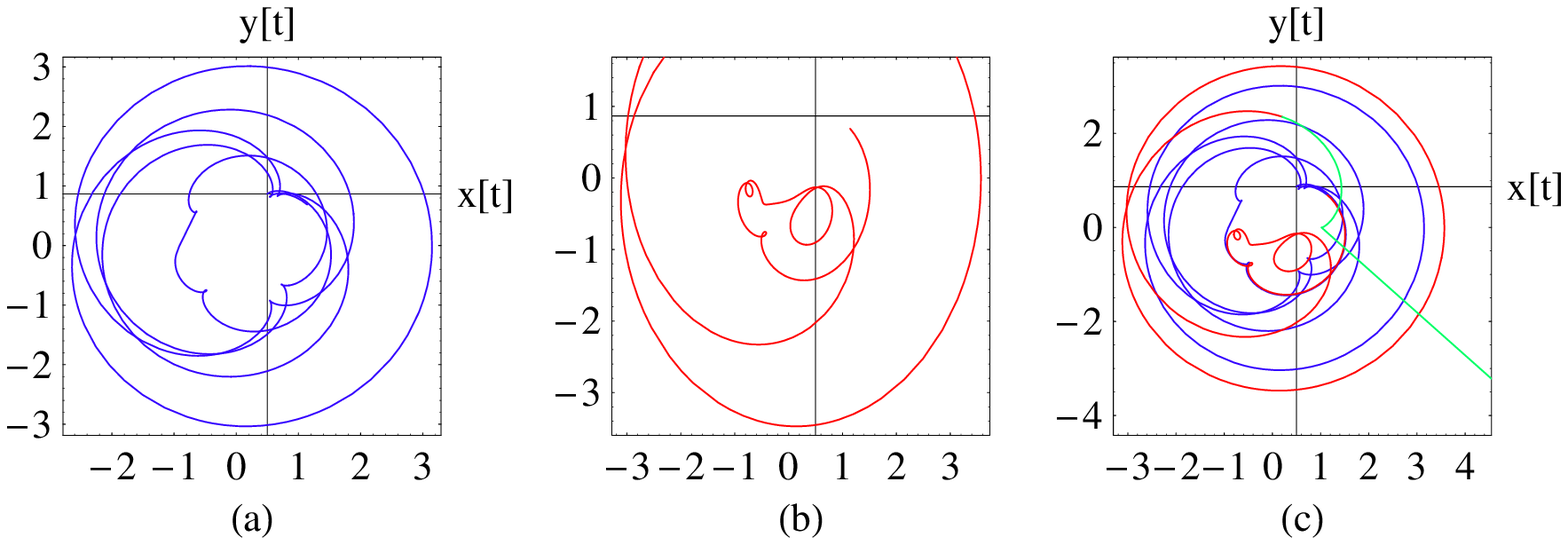}\\
\includegraphics[scale=.6]{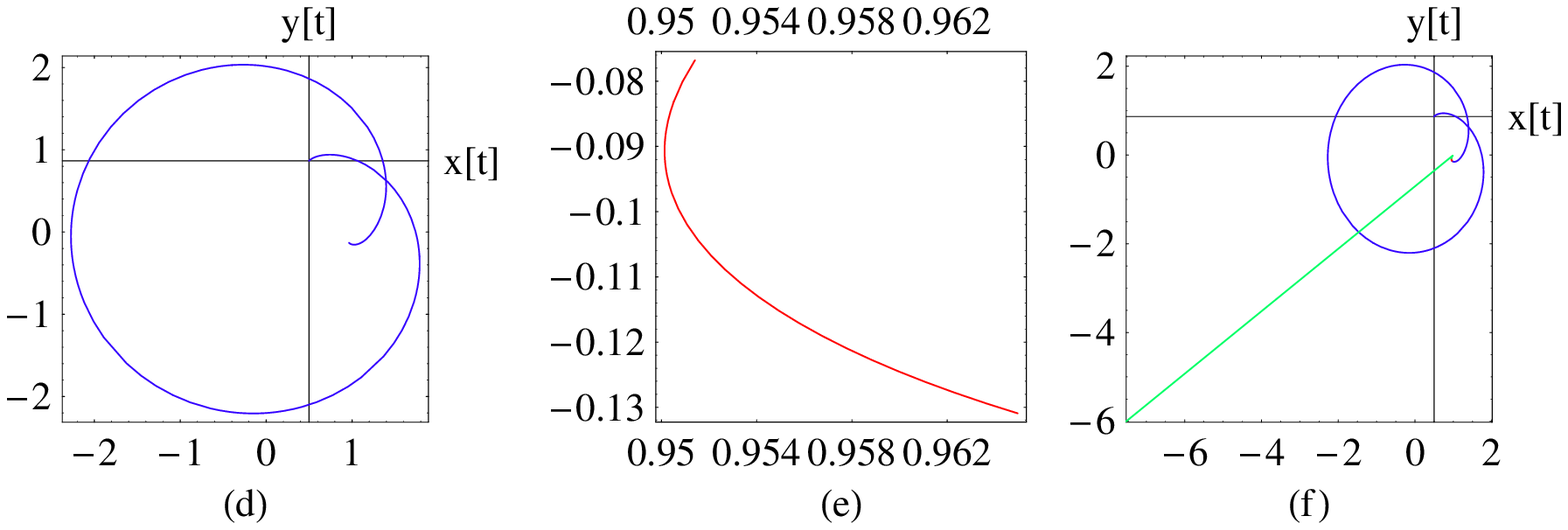}
\end{center}
\caption{Trajectory of  $L_4$  when  $q_1=0.75, M_b=0.25$, where frames(a-c) for  $A_2=0.25$ and  frames(d-f) for  $A_2=0.50$.}\label{fig:trjq75m25a}
\end{figure}
 
 When  $q_1=0.75, M_b=0.25$, figure  \ref{fig:trjq75m25a} depicts the trajectory for $L_4$  with  frames(a-c) for $A_2=0.25$  and  frames(d-f) for $A_2=0.50$. In frame(a) $0 \leq t \leq 50$, (b) $50 \leq t \leq 75$ and (c) $0 \leq t \leq 77$ while (d) $ 0 \leq t \leq 8.2$, (e) $8.2 \leq t \leq 8.3$ and (f)$0 \leq t \leq 9$. It is clear from frames(a-c) that the trajectory moves along approximately epicycloid path, when $t$ increases it departs form the  vicinity of $L_4$. The region of stability  shrinks and trajectory moves along  a single cusped epicycloid, then it departs far from the initial point. Hence oblateness effect is significant factor  to reducing the stability region.

\section{Lyapunov Characteristic Exponents(LCEs)}
\label{sec:lces}
It is well known that, if LCE$>0$ for some initial conditions which indicates the trajectory of initial condition is unstable. If LCE$=0$ for some values of initial conditions the orbit is neutrally stable and which corresponds to regular motion. If LCE$<0$, the corresponding orbit is asymptotically stable. 
 Now suppose $S$ be a 4-dimensional phase space such that \(S=\{X:X=[x(t),y(t),p_x(t),p_y(t)]^{Tran}\}\) , then the time evaluation of the orbit is governed by the equation 
\begin{equation}
\dot{X}=f(X)=\left[\frac{\partial{H}}{\partial{x}}\ \frac{\partial{H}}{\partial{y}}\ -\frac{\partial{H}}{\partial{p_x}}\ -\frac{\partial{H}}{\partial{p_y}}\right]^{tran}=J_4 D H X \label{eq:xt}\end{equation}
where $D=\frac{\partial{}}{\partial{X}}$ and 
 \begin{equation}
J_4=\begin{bmatrix}
0 &0 &1 &0 \\
-1&0&0&1\\
0&-1&0&0
\end{bmatrix}.
\label{eq:J}\end{equation}

The  dynamical system is described by the   Hamiltonian $H$ which depends on Jacobian constant and given by 
\begin{equation}
H= \frac{1}{2}\left(p_x^2+p_y^2\right)+n(yp_x-xp_y)-U(x,y)
\label{eq:H}
\end{equation} 
where
$p_x,p_y$ are the momenta coordinates given by
\[
\dot{p_x}=-\frac{\partial{H}}{\partial{p_x}},\ 
\dot{p_y}=-\frac{\partial{H}}{\partial{p_y}},
\]
\begin{eqnarray*}
 U(x,y)=\Omega-\frac{n^2(x^2+y^2)}{2},
\end{eqnarray*}

Consider $v=(\delta{x},\delta{y},\delta{p_x},\delta{p_y})$ be a deviation vector from initial condition $X(0)$ such that $||v_0||=1$.  Then the  variational equation  is given
\begin{equation}
\dot{v}=Df(X).v
\end{equation}
\begin{equation}
\mbox{or}\quad \begin{bmatrix}
\delta\dot{x}\\
\delta\dot{y}\\
\delta\dot{p_x}\\
\delta\dot{p_y}
\end{bmatrix}=\begin{bmatrix}
0 & n &1 & 0\\
-n& 0& 0&1\\
U^t_{xx}&U^t_{xy}&0&n\\
U^t_{yx}&U^t_{yy}&-n&0
\end{bmatrix}
\begin{bmatrix}
\delta{x}\\
\delta{y}\\
\delta{p_x}\\
\delta{p_y}\end{bmatrix},\label{eq:vareq}
\end{equation}
 where superscript $t$ over partial derivatives of $U$ indicates their  respective values at $t$ etc. Then the Lyapunov Characteristic Exponent is given by 
 \begin{equation}
  \lambda(v(t))=\lim_{t->\infty}\log\frac{||v(t)||}{||v(0||}.\label{eq:lcevt}
 \end{equation}
 
\begin{figure}
\includegraphics[scale=.5]{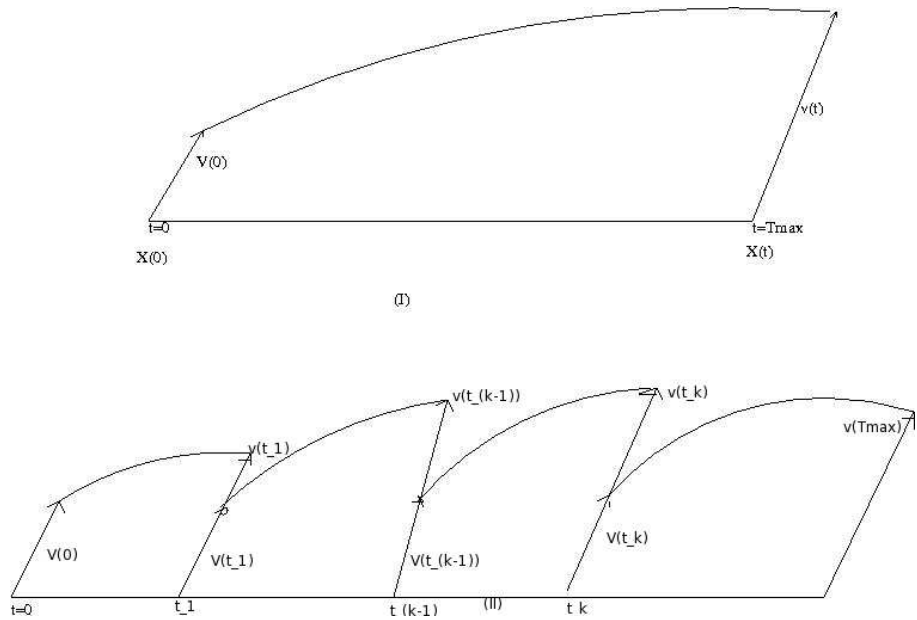}
\caption{In Plot (I) only  one step is used to obtain LCEs while in (II)  more that one normalization steps are used and $V(t_{k-1})$ denotes the unit deviation vector for all $k$.}\label{fig:lces}
\end{figure}

For numerical computation of LCEs we  use method presented in \cite{Skokos2010PhRvE..82c6704S} and \cite{Skokos2010LNP...790...63S}. To avoid overflow in numerical computation,  we partition  the closed interval $I=[t_0,Tmax]$  into $n_1$ sub intervals with time step $\Delta t$ and the time to run from $0$ to $Tmax$ i.e.
 \begin{equation}
 P(I)=\left\{0=t_0,t_1,t_2,t_3,\dots t_{k-1},t_k,\dots, t_{n_1}=Tmax\right\},
 \end{equation}
 then  equation (\ref{eq:lcevt}) can be written as 
 \begin{equation}
 \lambda(v(t))=\lim_{{n_1} t->\infty}\sum_{k=0}^{n_1}{\log \alpha(t_k)},\label{eq:lcePart}
 \end{equation}
where $\alpha(t_k)=||v(t_k)||$. 
To determine first order LCEs in next section, we will use   initial vector $X(0)= (0.499046, 0.866025, -0.866025, 0.499046)$ for  classical RTBP($q_1=1,A_2=0, M_b=0$) and  $X(0)=(0.337957, 0.81415, -0.954676, 0.39629)$ for modified RTBP( $q_1=0.75,A_2=0.25, M_b=0.25$). As in figure \ref{fig:lces},  at each step $v(t_k)$  will be  evaluated from (\ref{eq:xt},\ref{eq:vareq})  using $X(t_{k-1})$  and unit deviation vector $\hat{v}(t_{k-1})=V(t_{k-1})$(say). 

\subsection{First Order  LCEs}
\label{subsec:1lce}
\begin{figure}
\plottwo{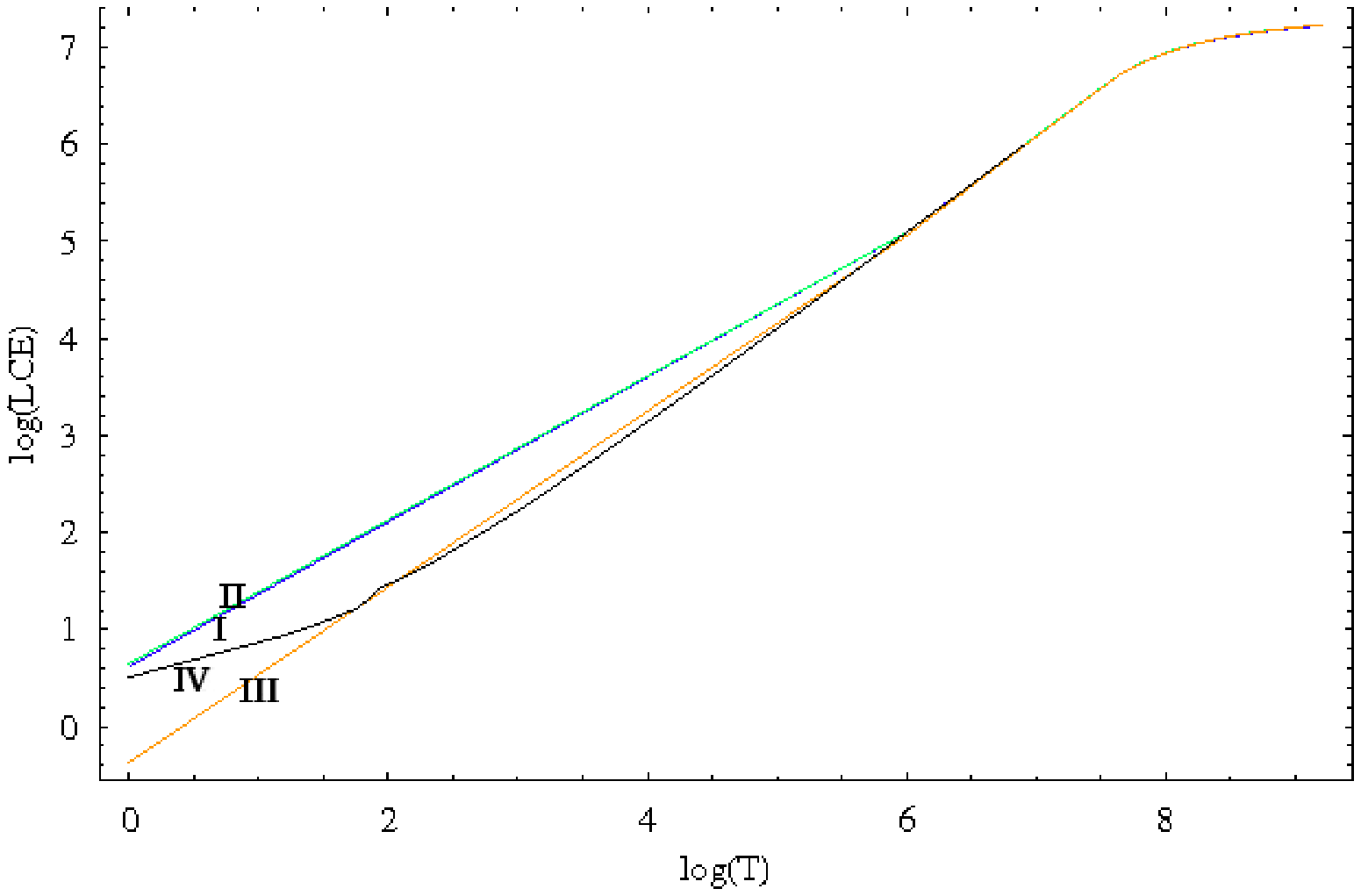}{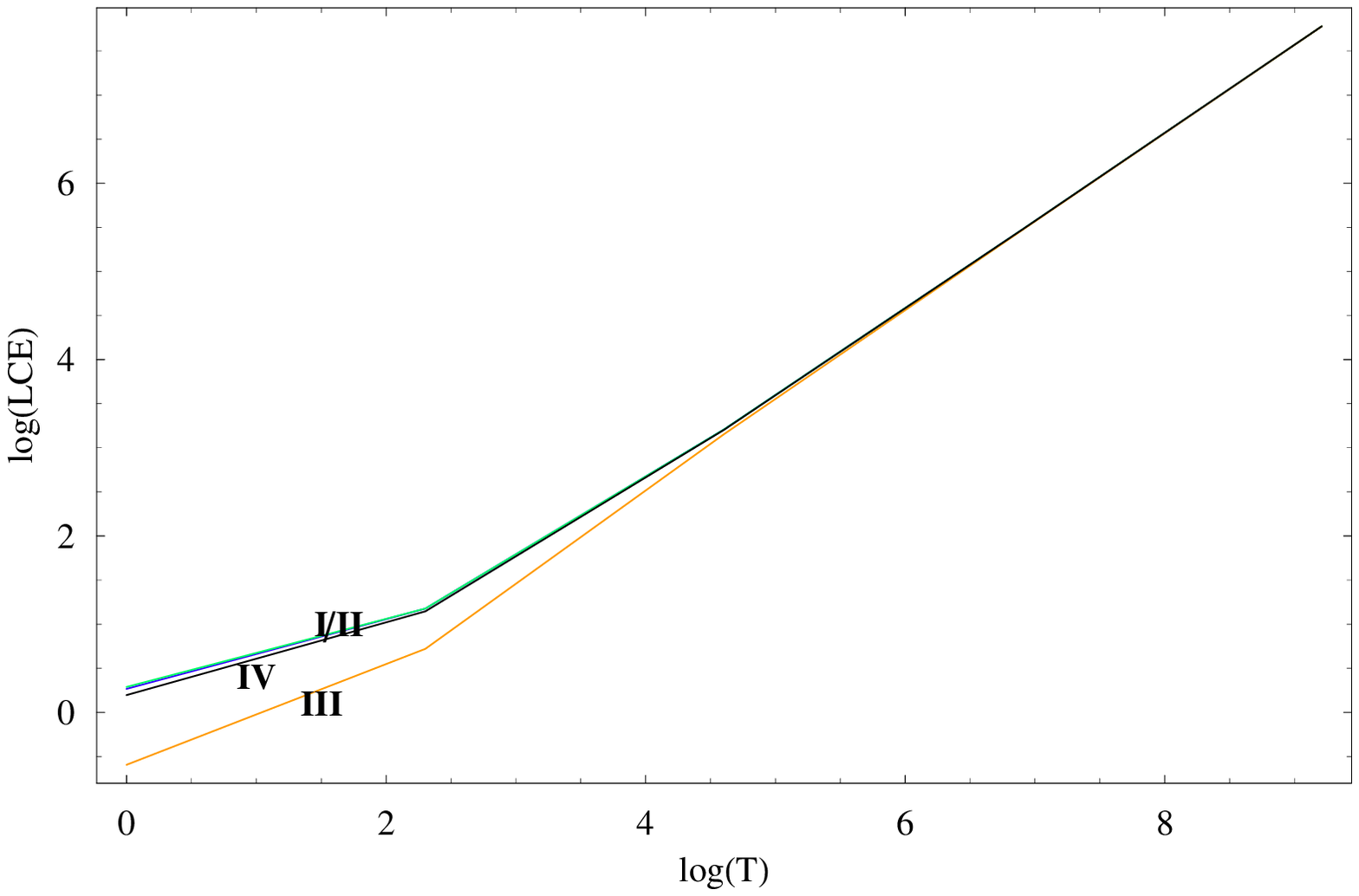}
\caption{LCE$(\lambda_1)$ when $q_1=0.75,A_2=0.25, M_b=0.25$ and $0 \leq t \leq 10000$; curve (I) $v_{11}=(1/2,1/2,1/2,1/2) $,(II):$v_{12}=(0,\frac{1}{\sqrt{3}},\frac{1}{\sqrt{3}},\frac{1}{\sqrt{3}})$, (III):$v_{13}=(0,0,\frac{1}{\sqrt{2}},\frac{1}{\sqrt{2}})$ and (IV):$v_{14}=(0,0,0,1).$\label{fig:qambL1}}
\end{figure}

\begin{figure}
\plottwo{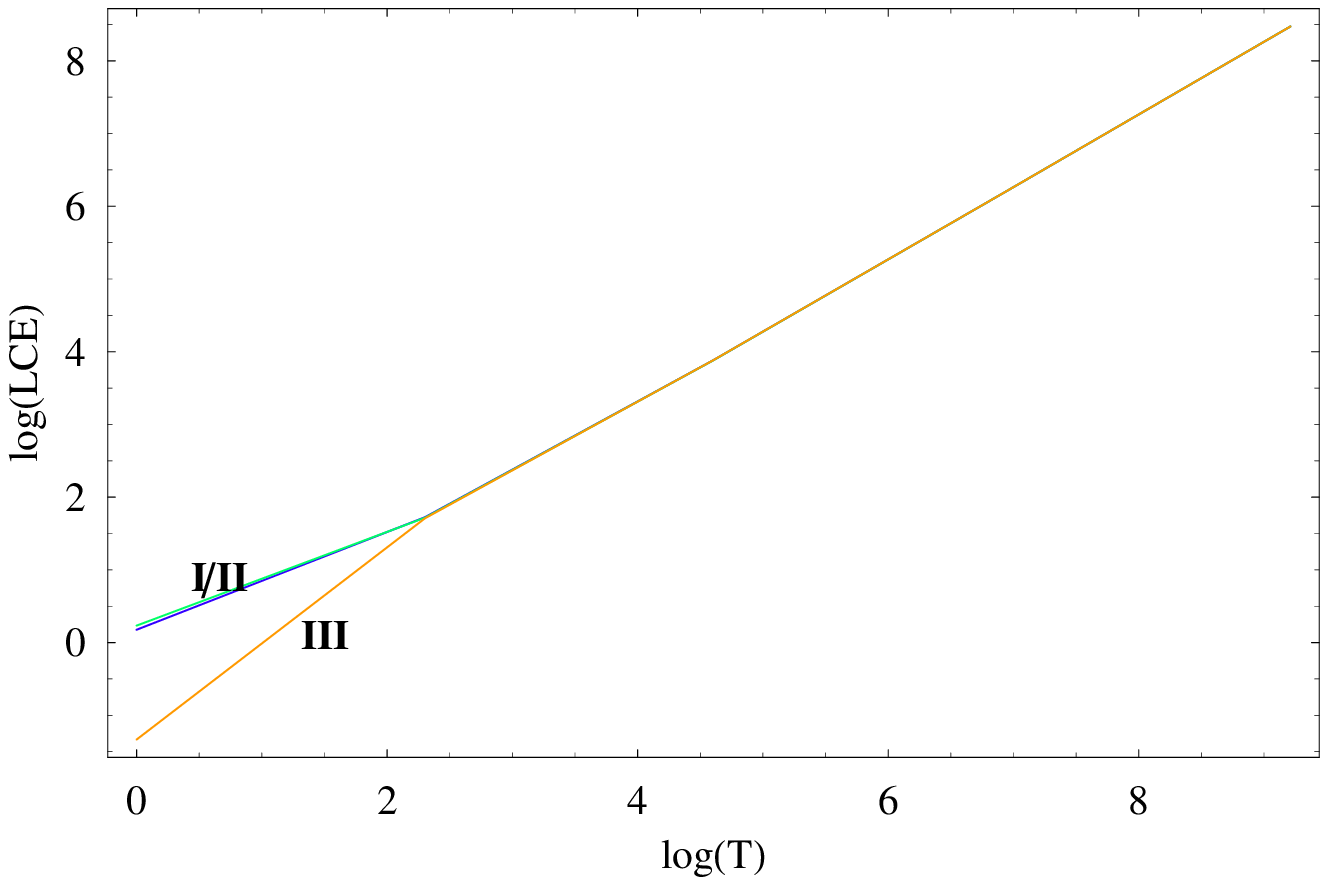}{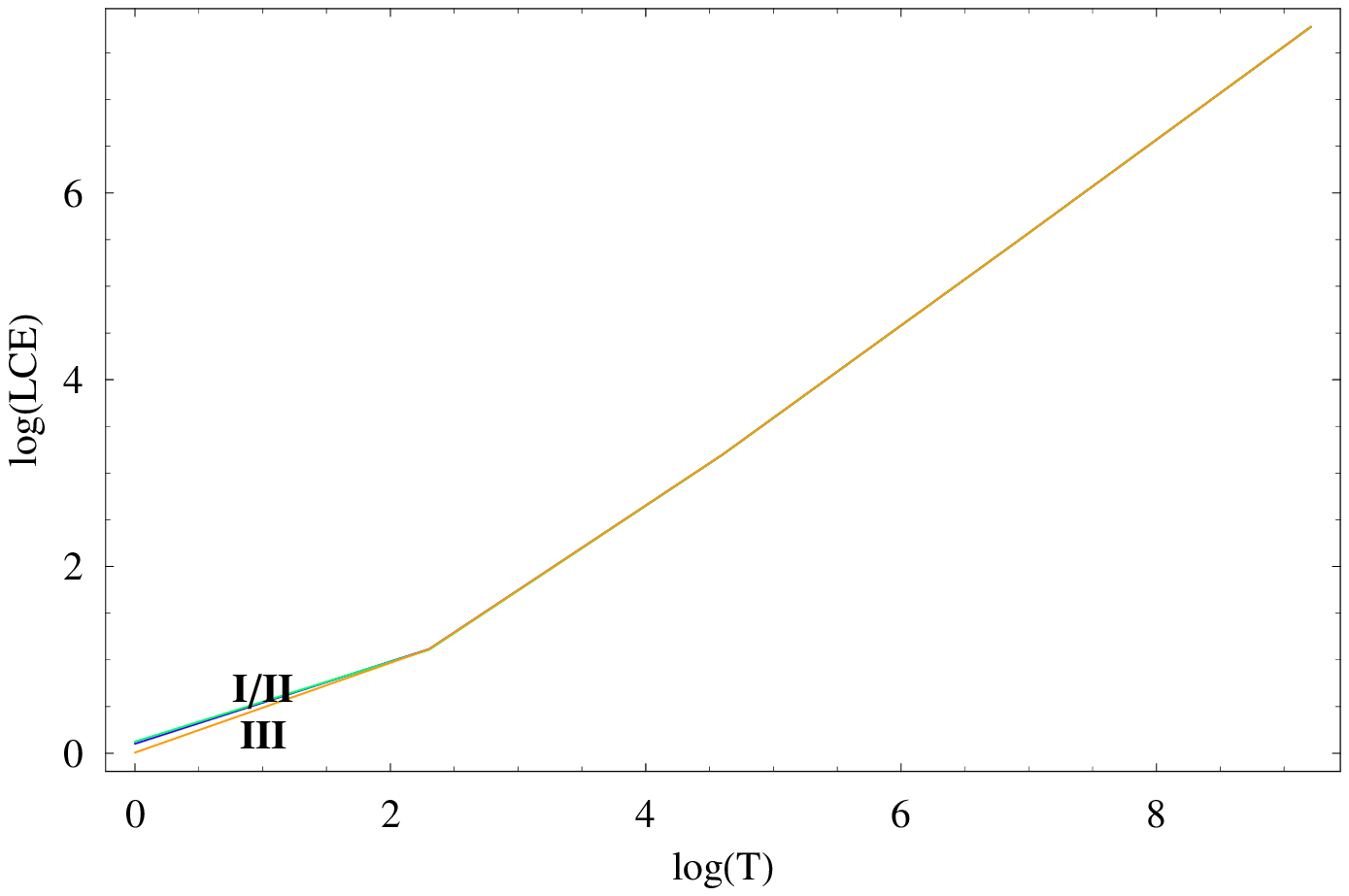}
\caption{LCE$(\lambda_2)$ when $q_1=0.75,A_2=0.25, M_b=0.25$ and $0\leq t\leq 10000$; curves (I) $v_{21}=(\frac{1}{\sqrt{3}},\frac{1}{\sqrt{3}},\frac{1}{\sqrt{3}},0) $,(II):$v_{22}=(0,\frac{1}{\sqrt{2}},\frac{1}{\sqrt{2}},0)$ and (III):$v_{23}=(0,0,1,0).$\label{fig:qambL2}}
\end{figure}

Now consider $R^4=LD_1$, $R^3=LD_2$, $R^2=LD_3$ and $R^1=LD_4$ spaces such that $LD_1\supset LD_2\supset LD_3\supset LD_4$. To find the first order LCE($\lambda_i),(i=1,2,3,4)$,  we choose  initial unit deviation vectors from $LD_1\backslash LD_2$: $v_{11}=(1/2,1/2,1/2,1/2) $, $v_{12}=(0,\frac{1}{\sqrt{3}},\frac{1}{\sqrt{3}},\frac{1}{\sqrt{3}})$, $v_{13}=(0,0,\frac{1}{\sqrt{2}},\frac{1}{\sqrt{2}})$, $v_{14}=(0,0,0,1)$.  The values  of LCEs are presented in  log-log plot figure \ref{fig:qambL1}  for $t= 0-10000$  when $q_1=0.75,A_2=0.25, M_b=0.25$ left panel corresponding to $\Delta t=1$ and right for $\Delta t=2$.  Initially the values of LCE($\lambda_1$) are different, they are   shown by curves(I)-(IV) correspond to   four vectors respectively,  but  when  $t$ increases they merge into a single curve. 
To obtain LCE($\lambda_2$), we choose  initial unit deviation vectors from  $LD_2\backslash LD_3$ such that  $v_{21}=(\frac{1}{\sqrt{3}},\frac{1}{\sqrt{3}},\frac{1}{\sqrt{3}},0)$, $v_{22}=(0,\frac{1}{\sqrt{2}},\frac{1}{\sqrt{2}},0)$,
$v_{23}=(0,0,1,0)$. Figure \ref{fig:qambL2}  shows LCE($\lambda_2$) when  $q_1=0.75,A_2=0.25, M_b=0.25$, where  left panel corresponds to $\Delta t=1$ and right for $\Delta t=2$.

\begin{figure}
\plottwo{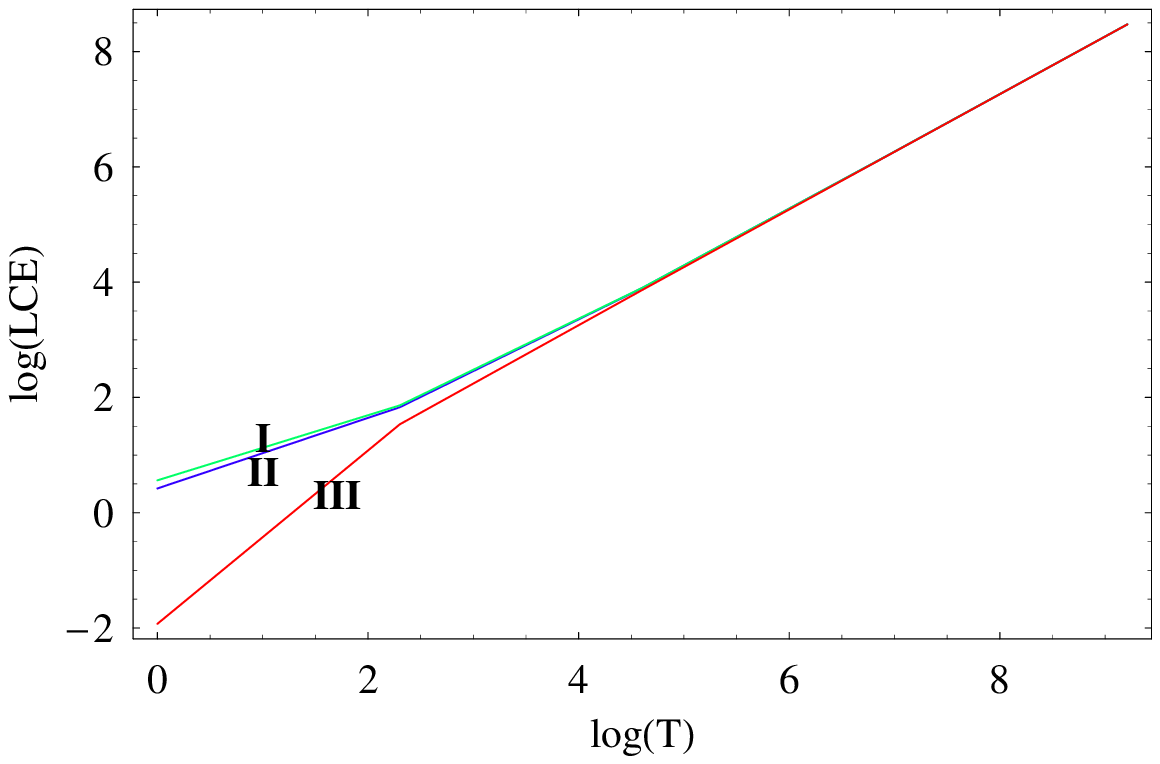}{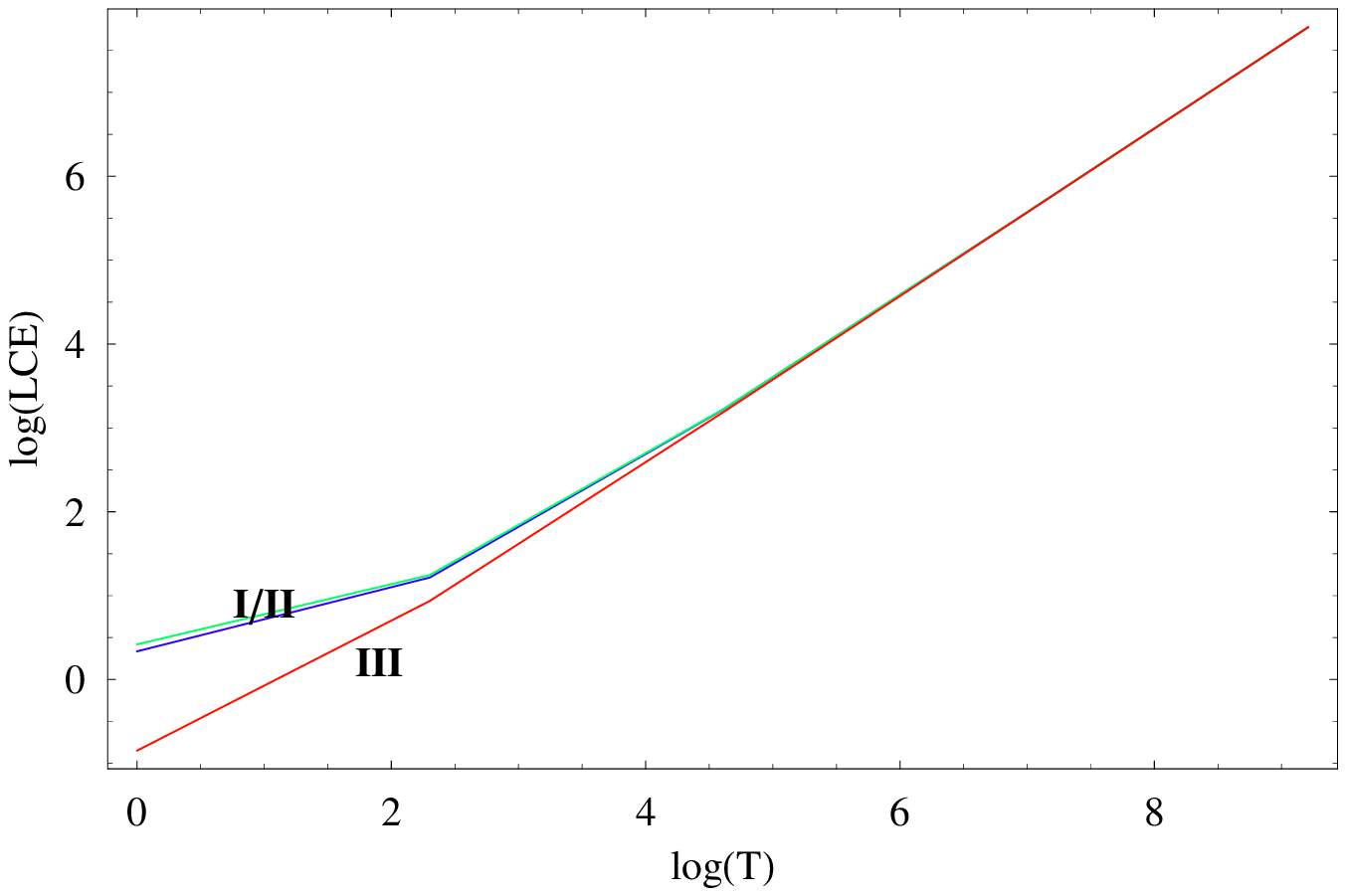}
\caption{LCEs when $0\leq t\leq 10000$,  $q_1=0.75,A_2=0.25$ and  $M_b=0.25$; where LCE$(\lambda_3)$:(I) $v_{31}=(\frac{1}{\sqrt{2}},\frac{1}{\sqrt{2}},0,0) $,(II):$v_{32}=(0,1,0,0)$ and  LCE$(\lambda_4)$(III):$v_{41}=(1,0,0,0).$\label{fig:qambL3L4}}
\end{figure}
\begin{figure}
\plottwo{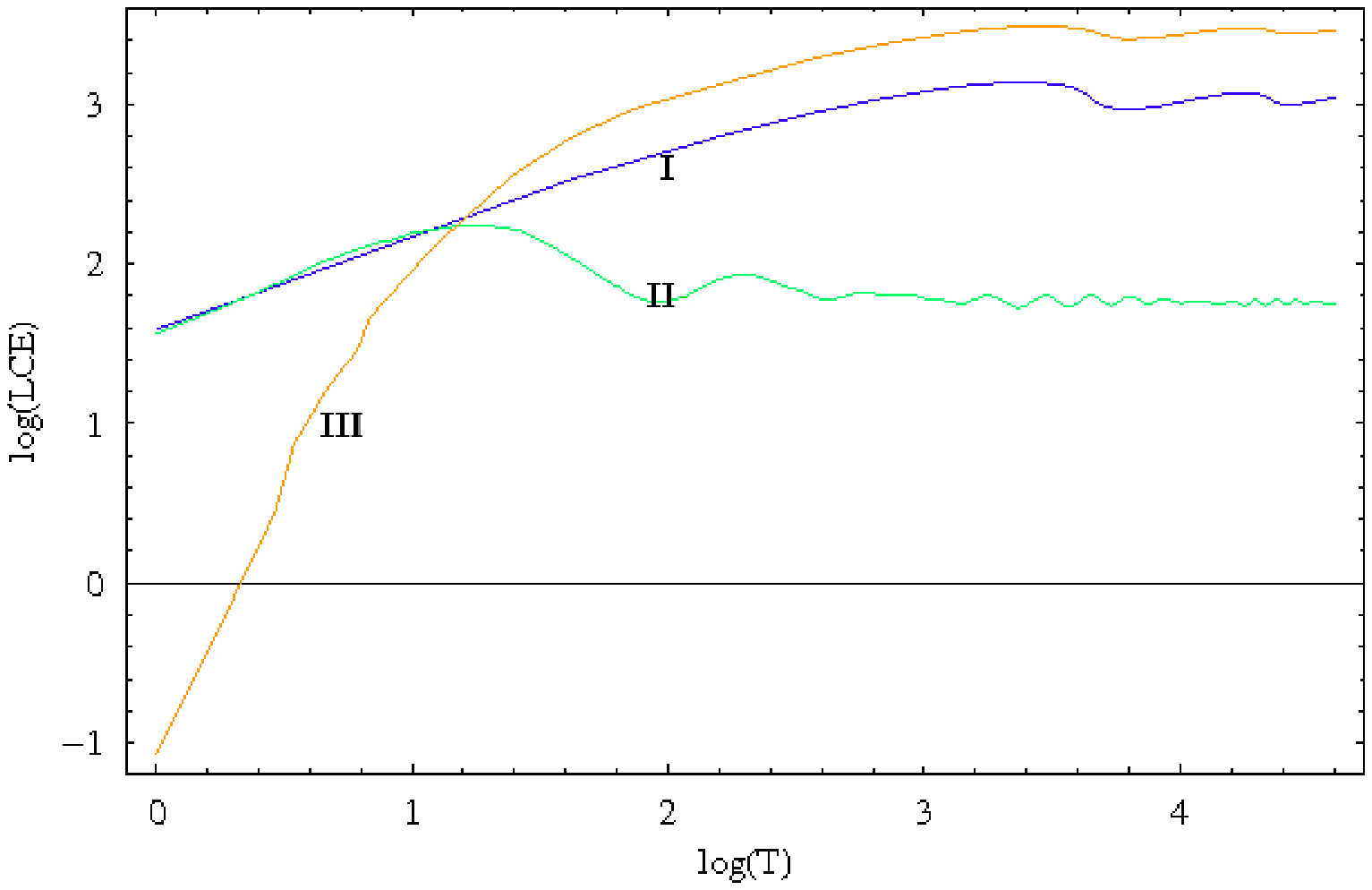}{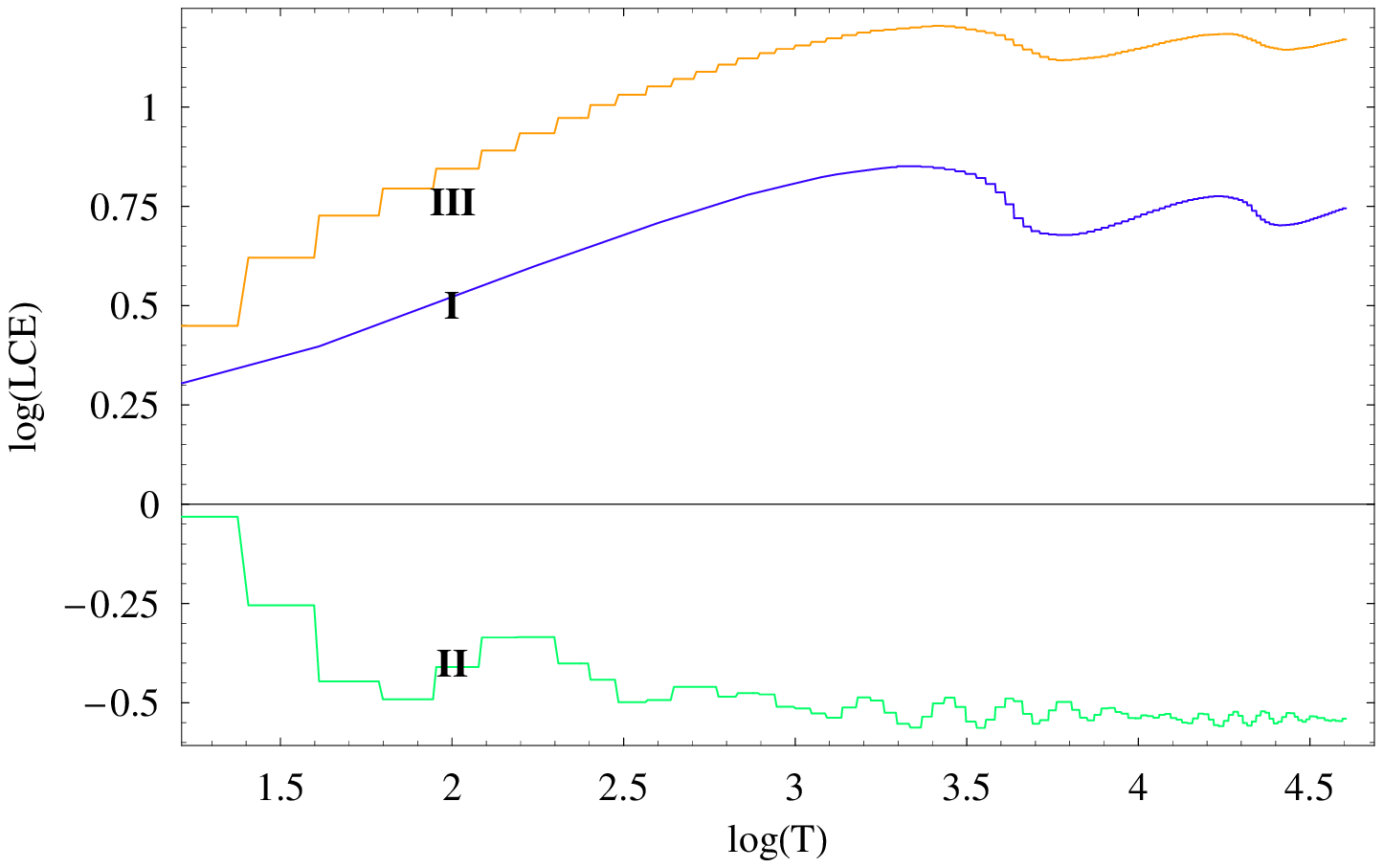}
\caption{LCE$(\lambda_2)$ when $q_1=1,A_2=0, M_b=0$ and $0\leq t\leq 100$; where (I):$v_{21}=(\frac{1}{\sqrt{3}},\frac{1}{\sqrt{3}},\frac{1}{\sqrt{3}},0) $,(II):$v_{22}=(0,\frac{1}{\sqrt{2}},\frac{1}{\sqrt{2}},0)$ and (III):$v_{23}=(0,0,1,0).$\label{fig:le100L2}}
\end{figure}

\begin{figure}
\plottwo{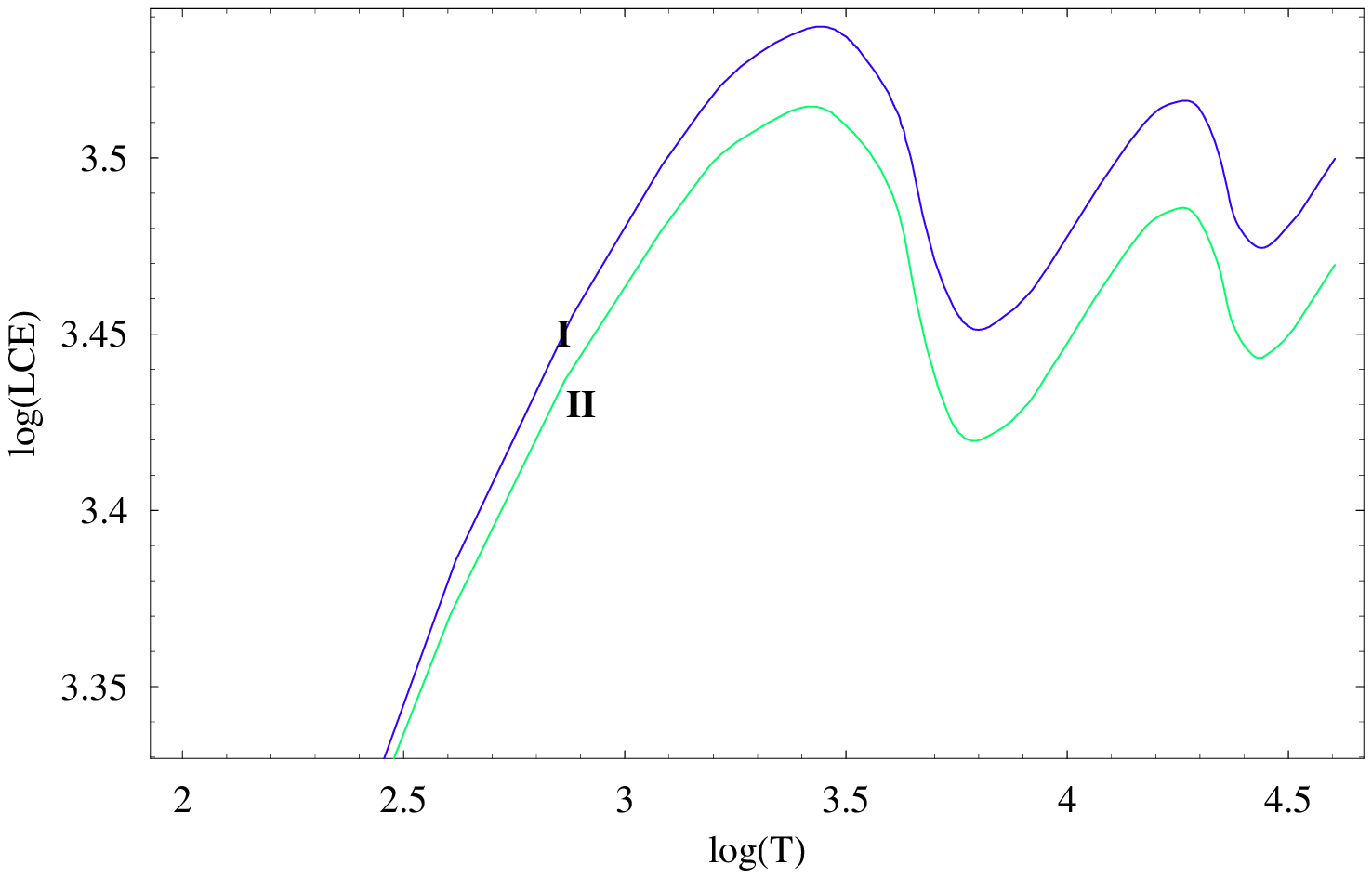}{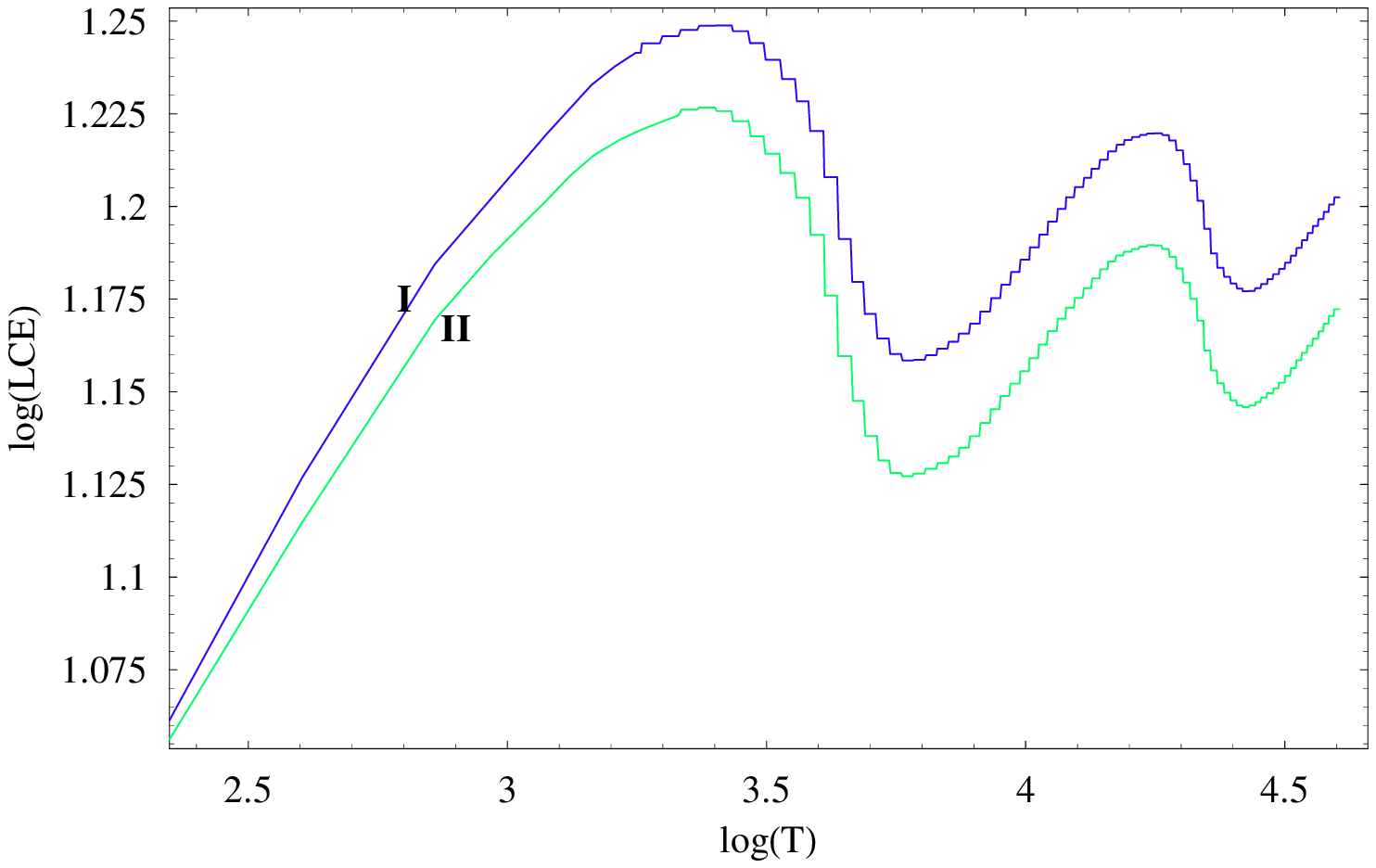}
\caption{LCE$(\lambda_3)$ when $q_1=1,A_2=0, M_b=0$ and $0\leq t\leq 100$; where  (I):$v_{31}=(\frac{1}{\sqrt{2}},\frac{1}{\sqrt{2}},0,0) $ and (II):$v_{32}=(0,1,0,0)$.}\label{fig:le100L3}
\end{figure}

Now for computation of LCE($\lambda_3$),  we choose initial unit deviation vectors from  $LD_3\backslash LD_4$. The results are presented in figure \ref{fig:qambL3L4} for $q_1=0.75,A_2=0.25, M_b=0.25$ and $0\leq t\leq 10000$ with left frame for  $\Delta t=1$ and right  for $\Delta t=2$. In  figures \ref{fig:le100L2} and \ref{fig:le100L3}, curves are   plotted  when  $q_1=1, A_2=0,M_b=0$,  where  left panel corresponding to $\Delta t=0.1$ and right for $\Delta t=1$. In  figure \ref{fig:le100L2}, curves are labeled as(I) $v_{21}=(\frac{1}{\sqrt{3}},\frac{1}{\sqrt{3}},\frac{1}{\sqrt{3}},0) $,(II):$v_{22}=(0,\frac{1}{\sqrt{2}},\frac{1}{\sqrt{2}},0)$, (III):$v_{23}=(0,0,1,0)$ and in figure  \ref{fig:le100L3}, curves are plotted for $0\leq t\leq 100$, where (I) $v_{31}=(\frac{1}{\sqrt{2}},\frac{1}{\sqrt{2}},0,0) $,(II):$v_{32}=(0,1,0,0)$. The curves are in wave form with decreasing amplitudes which tend to zero at infinity and  curves become constant. 

\begin{figure}
\plotone{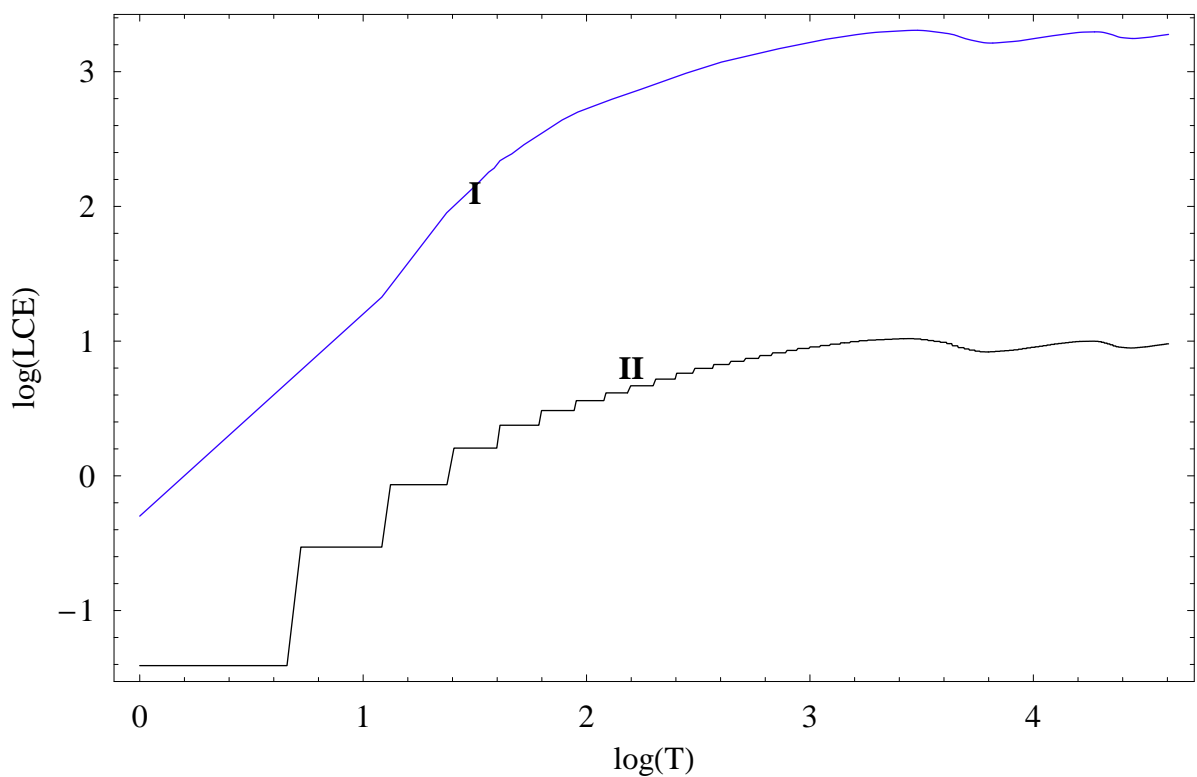}
\caption{LCE$(\lambda_4)$ when $q_1=1,A_2=0.0, M_b=0.0$ and $v_{41}=(1,0,0,0) $ where  curve(I) represents the renormalization time step $\Delta t=0.1$  and (II) for $\Delta t=1$.}\label{fig:le100L4}
\end{figure}

To determine LCE$(\lambda_4)$, we choose $v_{41}=(1,0,0,0)$ from $LD_4$ . The corresponding LCE is  shown by curve (III) in  figure  \ref{fig:qambL3L4}:$q_1=0.75,A_2=0.25, M_b=0.25$ with left frame for  $\Delta t=1$ and right  for $\Delta t=2$.  In  figure  \ref{fig:le100L4}, we consider $q_1=1,A_2=0.0, M_b=0.0$ in which  curve(I) represents renormalization time step $\Delta t=0.1$  and (II) for $\Delta t=1$. It can be seen that (I) is a smooth  curve and (II) is initially  stepped curve but    both  curves are  initially increasing in nature and after certain time they become constant.
The  LCEs$(\lambda_i),  (i=1,2,3,4)$ are presented in Table \ref{tab:MaxTforL2}  for   initial point   $X(0)=(0.337957, 0.81415, -0.954676, 0.39629)$ and $q_1=0.75,A_2=0.25,M_b=0.25.$ It is clear from figures and Table  that all  first order LCEs  are positive for various set values of parameters and  renormalization time steps.  This shows that  the  present   dynamical system  is  stochastic. It is also noticed that if $Tmax$ is not  very large the LCEs depend on the choice of the renormalization time step as well as initial deviation vectors while  if $Tmax$ is very large  then  LCEs depend on renormalization time  step only. 

\clearpage
\begin{deluxetable}{rrrrrrrrrrr}
\tabletypesize{\scriptsize}
\rotate
\tablecaption{First order  LCEs for initial point $X(0)=(0.337957, 0.81415, -0.954676, 0.39629)$ when $q_1=0.75,A_2=0.25,M_b=0.25.$\label{tab:MaxTforL2}}
\tablewidth{0pt}
\tablehead{
\colhead{$\log t $} & \colhead{$\log{\lambda_1(v_{11})}$} & \colhead{$\log{\lambda_1(v_{12})}$} & \colhead{$\log{\lambda_1(v_{13})}$} & \colhead{$\log{\lambda_1(v_{14})}$} &
\colhead{$\log{\lambda_2(v_{21})}$} & \colhead{$\log{\lambda_2(v_{22})}$} & \colhead{$\log{\lambda_2(v_{23})}$} &
\colhead{$\log{\lambda_3(v_{31})}$} & \colhead{$\log{\lambda_3(v_{32})}$}&\colhead{$\log{\lambda_4(v_{41})}$}
}
\startdata
&&&&&$\Delta t= 1$&&&&&\\\hline
$\log 10^0$  &0.403836&0.446978 &-0.474575&0.182934&0.176488&0.233361&-1.33349&0.420617&0.561358&-1.92815\\
$\log 10^1$  &1.78922& 1.79078 &1.31139&1.75423&1.72431&1.71777&1.70948&1.83046&1.86196&1.53422\\
$\log 10^2$ & 3.89211& 3.89209&3.83842&3.88798&3.88433&3.88314&3.88489&3.89796&3.90179&3.8656\\
$\log 10^3$ & 6.17181& 6.17188&6.16639&6.17139&6.17102&6.17098&6.1711&6.17242&6.17281&6.16922\\
$\log 10^4$ & 8.47207&8.47216 &8.47153&8.47203&8.47199&8.47208&8.472&8.47213&8.47217&8.47188\\\hline
&&&&&$\Delta t= 2$&&&&&\\\hline
$\log 10^0$  &0.266298& 0.287879 &-0.594977&0.195472&0.101962&0.124369&0.00722875&0.335259&0.418618&-0.849014\\
$\log 10^1$  &1.17546& 1.17649 &0.719618&1.1448&1.11437&1.10752&1.11319&1.21633&1.24543&0.936813\\
$\log 10^2$ &3.2087& 3.20867&3.15533&3.20463&3.20098&3.19979&3.20168&3.21453&3.21832&3.18239\\
$\log 10^3$ & 5.47966& 5.47973&5.47424&5.47924&5.47887&5.47883&5.47895&5.48027&5.48066&5.47707\\
$\log 10^4$ &7.77903&7.77911 &7.77848&7.77898&7.77895&7.77903&7.77895&7.77909&7.77913&7.77884\\\hline
&&&&&$\Delta t= 10$&&&&&\\\hline
$\log 10^1$  &0.030892& 0.030311 &-0.25883&0.011177&-0.0066539&-0.013192&-0.002721&0.0587713&0.076341&-0.10142\\
$\log 10^2$ &1.67368& 1.67363&1.6238&1.66989&1.66652&1.66538&1.66726&1.67913&1.68263&1.64948\\
$\log 10^3$ & 3.87816& 3.87824&3.87278&3.87775&3.87738&3.87734&3.87746&3.87877&3.87915&3.87559\\
$\log 10^4$ &6.17039&6.17047 &6.16985&6.17035&6.17031&6.17039&6.17032&6.17045&6.17049&6.1702\\\hline
&&&&&$\Delta t= 100$&&&&&\\\hline
$\log 10^2$ &-0.0374842&-0.0374685&-0.064786&-0.039577&-0.0414376&-0.042026&-0.0410299&-0.0344616&-0.0325176&-0.0507718\\
$\log 10^3$ &1.66081& 1.66088&1.65587&1.66043&1.66009&1.66006&1.66016&1.66136&1.66172&1.65846\\
$\log 10^4$ &3.87675&3.87684&3.87621&3.87671&3.87667&3.87676&3.87668&3.87681&3.87685&3.87657\\\hline
&&&&&$\Delta t= 1000$&&&&&\\\hline
$\log 10^3$ &1.66081& 1.66088&1.65587&1.66043&1.66009&1.66006&1.66016&1.66136&1.66172&1.65846\\
$\log 10^4$ &3.87675&3.87684&3.87621&3.87671&3.87667&3.87676&3.87668&3.87681&3.87685&3.87657\\\hline
&&&&&$\Delta t= 10000$&&&&&\\\hline
$\log 10^4$ &-0.0452992&-0.0452123&-0.0455707&-0.0453203&-0.045339&-0.0452492&-0.0453349&-0.0452687&-0.045249&-0.0453539\\
\hline
\enddata
\tablecomments{The values of LCEs  depend  on time step for normalization more  than   the initial deviation vector.}
\end{deluxetable}
\clearpage
\section{Stability  of  $L_4$}
\label{sec:stbL4}
\begin{figure}
    \includegraphics[scale=.6]{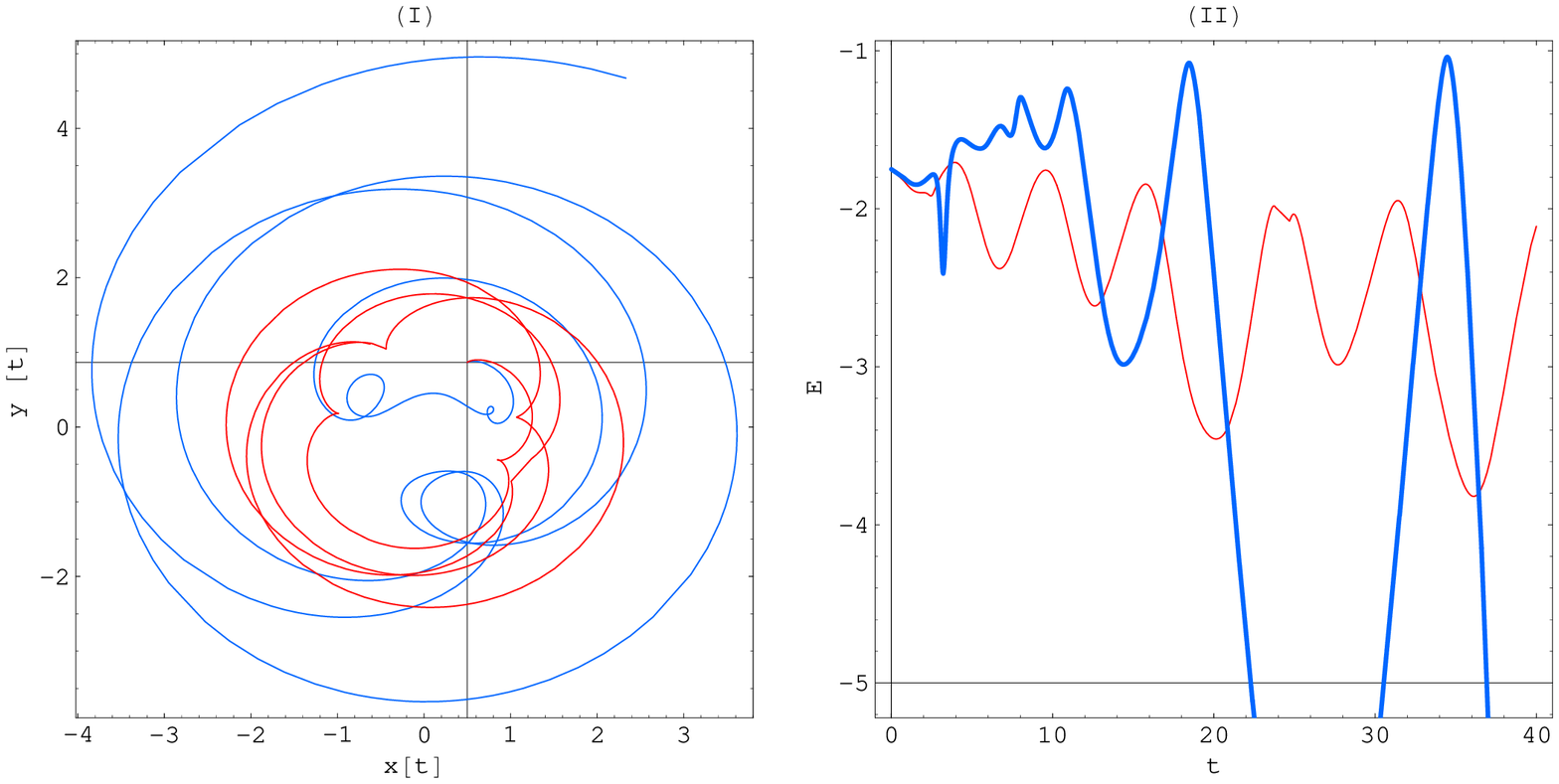}\\\includegraphics[scale=.6]{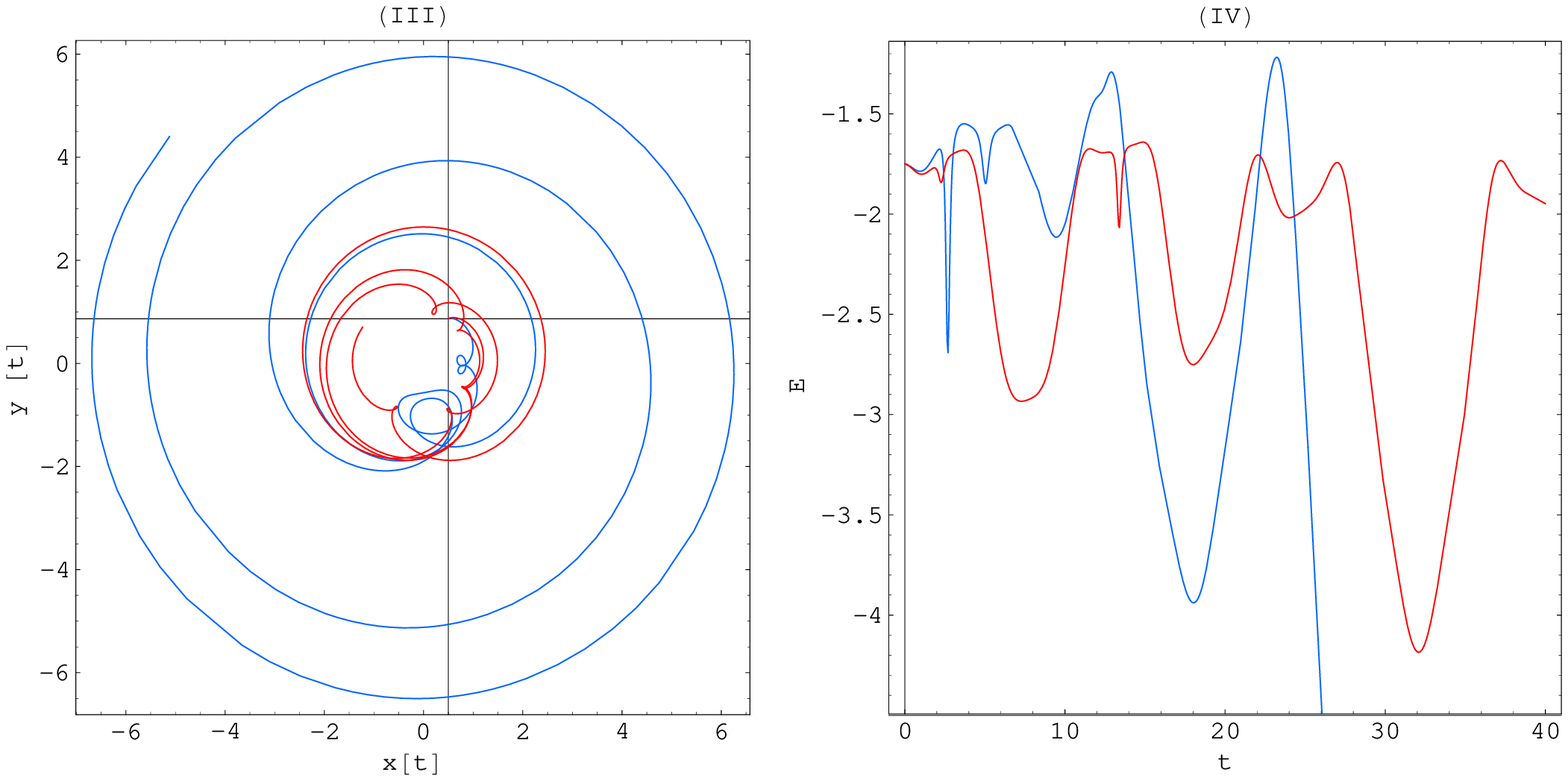}
   \caption{Stability of $L_4$ when $A_2=0.0$ with panels (I\&II):$q_1=0.75$ and (III\&IV):$q_1=0.50$ in which blue curves correspond to $M_b=0.25$ and red  for $M_b=0.50$.}
   \label{fig:stb}
\end{figure}

Now we suppose that   the coordinates $(x_1, y_1)$ of $L_4$ are
initially perturbed by changing \(x(0) = x_1+\epsilon \cos(\phi),
y(0) = y_1+\epsilon\sin(\phi)\) where \( \phi
=\arctan\left(\frac{y(0)-y_1}{x(0)-x_1}\right)\in  (0, 2\pi), 0 \leq
\epsilon < 1\);   $\phi$
indicates the direction of the initial position vector in the local
frame. For simplicity, it is supposed that  $\epsilon = 0.001$ and  $\phi= \frac{\pi}{4}$. We  solved   (\ref{eq:Omegax}, \ref{eq:Omegay}) numerically  using above perturbed initial point  and plotted  figure \ref{fig:stb}  when $A_2 = 0.0$,  which shows that  the orbit  of test particle and its energy constant. When $ q_1 = 0.75$ we have  panels(I\&II) and  $ q_1 = 0.50$ then  (III\&IV) in which (I\&III)   describe  the  trajectory and (II\&IV) correspond to energy integral $E$.  It is clear from the orbit that  initially trajectory  moves in epicycloid  path  described by (\ref{eq:epitx}, \ref{eq:epity})  without deviating far from  $L_4$ and energy constant remains negative; but  after a   certain time  it moves   spirally outward from the region and energy constant becomes positive. Here blue curves  represent $M_b = 0.25$ and red curves correspond to $M_b = 0.50$.

\begin{figure}
      \includegraphics[scale=.6]{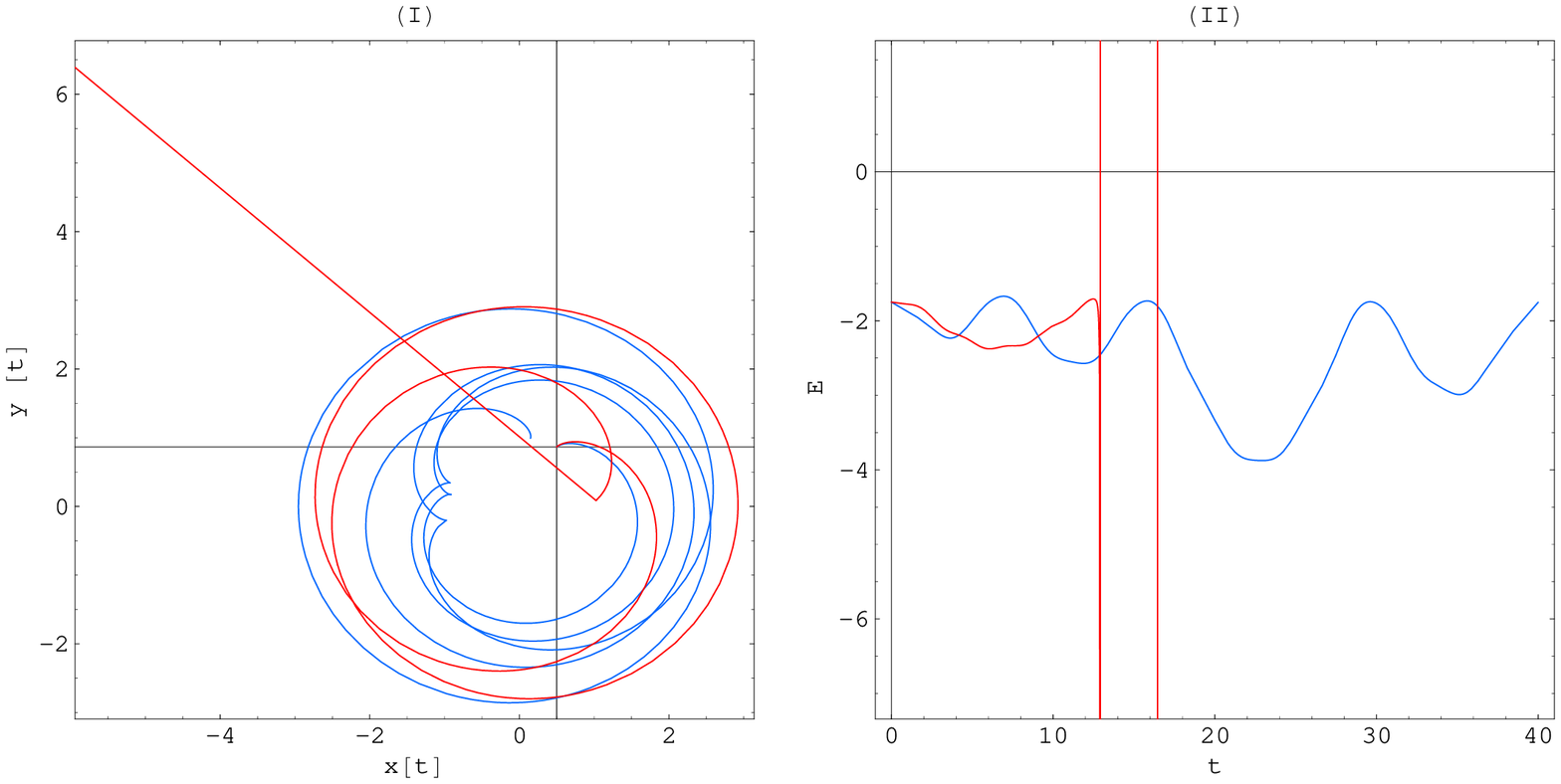}\\\includegraphics[scale=.6]{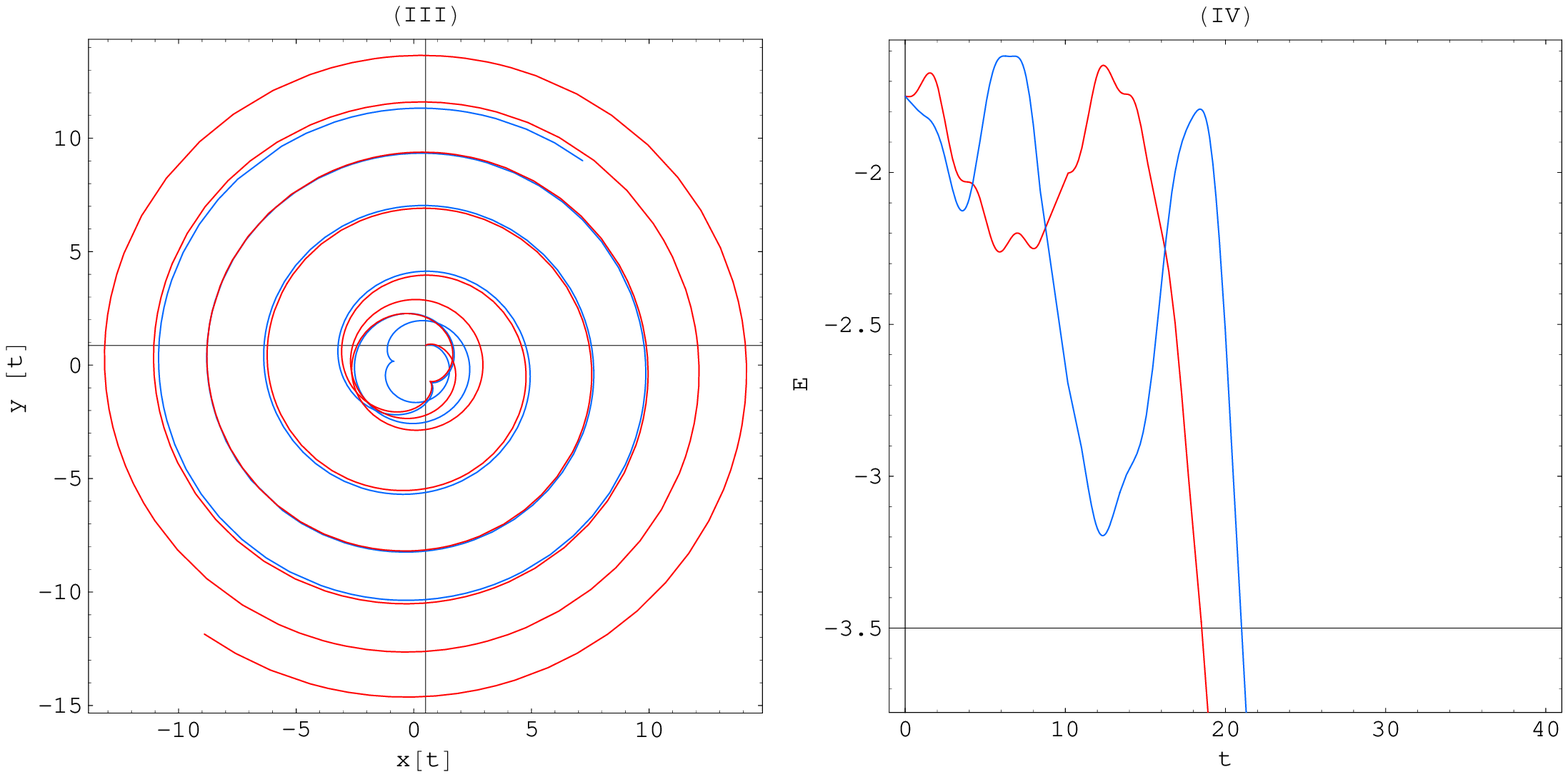}
   \caption{Effect of oblateness on the stability of $L_4$ when $M_b=0.25$ panels (I\&II)$q_1=0.75$  (III\&IV) $q_1=0.50$  in which blue solid curves for $A_2=0.25$, red curves for $A_2=0.50$.}
   \label{fig:stbOBLT}
\end{figure}

The effect of oblateness of the second primary is shown in figure
\ref{fig:stbOBLT} when $M_b=0.25$, where (I\&II) correspond to $q_1=0.75$ and  (III\&IV) for $q_1=0.50$. Panels (I\& III) show the  trajectory of perturbed point $L_4$ and (II\&IV) describe the energy integral 
of that point. The blue  curves  correspond to $A_2=0.25$ and red
 for $A_2=0.50$. The trajectory of perturbed point follows the path described by epitrochoid (\ref{eq:epitx},\ref{eq:epity}), as  time increases  it  moves  spirally outward from the vicinity of $L_4$.   It is seen  that the oblateness  is  a significant effect  on the trajectory and the stability of $L_4$. When $A_2=0.0$
the $L_4$ is asymptotically stable  for the value of $t$ which lies
within a certain interval. But if oblate effect of second primary is
present($A_2\neq 0$), the stability region of $L_4$ disappears for large values of $A_2$.

 From relation  $A_{2}=\frac{r^{2}_{e}-r^{2}_{p}}{5r^{2}}$ we obtain  \(\frac{r_e}{r}=\sqrt{5 A_2+\frac{r_p}{r}}\). This shows that if $A_2$ increases means the ratio $\frac{r_e}{r} $ increases consequently $ \frac{r_e}{r_1}$ increases. Then form   \cite{ryabov2006elementary},  it is  found that  as the attracting particle recedes  that is  the ratio  $ \frac{r_e}{r_1}$ diminishes, the difference between the attraction of spheroid and that of a sphere will decrease and, if $r$(or $r_1$) is very large in comparison with $r_e$, the spheroid will exert a force that practically coincided with that of a sphere.  If $A_2=0.5$ i.e. very large value then $r_e>r$(hypothetically) then second primary becomes a thin flat disk. In this case the  both primaries have no separate gravitational attraction so  they act like a single body and its sphere of influence is common with very large radius, that attracts perturbed point.  Since perturbed point $L_4$ is supposed  in the equatorial plane of second primary, so the attraction of the  oblate spheroid(equatorial bulge) upon $L_4$ at a given distance from the centre  of primary is greater than that of a sphere of equal mass($A_2=0.0$) which has been   proved by  \cite{moulton}. The effect  of oblateness can be seen in frame(c) of figure \ref{fig:trjq75ma0} and in frame(I) of  figure \ref{fig:stbOBLT}, where attraction of the equatorial bulge of the  second primary increases with $A_2$. Hence from above discussion we can say that if $A_2=0.50$(hypothetically),  the trajectory suddenly moves from the vicinity of $L_4$ as the time increase.  
\section{Conclusion}
\label{sec:con}
We have obtained the trajectories of $L_4$ and its  perturbed point,  for various set values of parameters. It is found that the trajectories  move along the epicycloid path upto  a certain time then  they  move spirally outward  from the vicinity of the point. From the first order Lyapunov Characteristic Exponents(LCEs), we have seen  that the behaviours of trajectories are stochastic. It is also found that the radiation pressure, mass of  the belt and oblateness  are  significant effects,  they reduce the stability region and increase the stochasticity in the system. It is also found that  if $A_2=0.50$(hypothetically),  the trajectory suddenly moves from the vicinity of $L_4$ as the time increase. 

\acknowledgments{I am thankful to the Department of Science \& Technology Govt. of India  for  providing financial support through SERC-Fast Track Scheme for Young Scientist in Physical Sciences (DO.No.SR/FTP/PS-121/2009,dated $14^{th}$ May 2010). I am also thankful to the Indian School of Mines, Dhanbad (India) for providing financial support through  Minor Research Project (No.2010/MRP/AM/04/Acad. dated $30^{th}$ June 2010.)}
\bibliographystyle{elsarticle-harv}

\end{document}